\documentclass[
aps,%
12pt,%
final,%
notitlepage,%
oneside,%
onecolumn,%
nobibnotes,%
nofootinbib,%
superscriptaddress,%
noshowpacs]%
{revtex4}%
\usepackage{geometry}
\usepackage{graphics}

\begin{document}
\title{Production and decay of charmed baryons: spectra of muons and 
asymmetry between $\mu^+$ and $\mu^-$}
\author{\firstname{O.~I.} \surname{Piskounova}}
\email[E-mail address: ]{piskoun@sci.lebedev.ru}
\affiliation{Lebedev Physics Institute, Russian Academy 
             of Sciences, Moscow }
\author{\firstname{N.~V.} \surname{Nikitin}}
\email[E-mail address: ]{nik679@monet.npi.msu.su}
\affiliation{SINP Moscow State University, Russia}

\begin{abstract}
The calculation of muon spectra from the decay of 
$\Lambda_c$ baryons was carried out on the basis of 
the description of recent data on charmed-baryon 
production in hadronic interactions. Data are described 
in the framework of Quark--Gluon String Model that 
allowes us to consider primary proton interactions of 
arbitrary high energy. MC code was built for charmed-baryon 
semileptonic decay in order to obtain the kinematical 
characteristics of resulting particles. It is predicted that
the charge asymmetry between energy spectra of $\mu^+$ and $\mu^-$
in laboratory system is clearly seen as the consequence of asymmetry 
between the spectra of charmed baryons and antibaryons.This 
extension of QGS Model can be useful to correct the 
calculations of muon and neutrino spectra in astrophysics.   
\end{abstract}

\maketitle

\section*{Introduction}
Charmed-particle decay is an important source of atmospheric muons and neutrinos \cite{cosmic_muons}. We know the characteristics of charmed-hadron production, investigated in the recent years by many fixed-target experiments in accelerators \cite{accelerator_charm}. 
Model of Quark--Gluon Strings (QGSM) \cite{general_QGSM} that has been built last two decades 
can reproduce the energy ditributions of charmed baryons and mesons in hadronic collisions 
up to very high energies. 
These spectra have an interesting behavior in the central region of $x_{{\rm F}}$: the asymmetry 
between particle and antiparticle distributions shows a narrow dip for charmed baryons,
otherwise the charmed meson asymmetry goes down slowly and reaches a nonzero value in the 
central $x_{{\rm F}}$ region, that contradicts the basical QCD theory. Practically it is not
possible to reproduce nonzero asymmetry in perturbative QCD approach due to the equal rate
of $c$ and $\bar{c}$ production in perturbative gluon fusion process. These phenomena have found
an explanation in recent calculations in the framework of QGSM \cite{previous_QGSM}. It seems 
interesting to investigate what sort of peculiarities should be produced by this dip
in the spectra of muons and neutrino. 

\section{Production of charmed baryons in $pp$ interactions and QGSM} 

The inclusive-production cross section of hadrons of type $H$ is written as
a sum over $n$-Pomeron cylinder diagrams: 

\begin{equation}
f_{1}=x \frac{d \sigma^{H}}{dx}(s,x)= \int E \frac{d^{3}\sigma^{H}}
{d^{3}p}d^{2}p_{\bot}=\sum_{n=0}^{\infty}\sigma_{n}(s) \varphi_{n}^{H}(s,x).
\end{equation}

Here, the function  $\varphi_{n}^{H}(s,x)$ is a particle distribution in the
configuration of $n$-cut cylinders and $\sigma_{n}$ is the probability of
this process. The cross sections $\sigma_{n}$ depend  on the
parameter of the supercritical Pomeron $\Delta_{{\rm P}}$, which is equal in our model
to 0.12 \cite{general_QGSM}. 
In the case of $\Lambda_c$ production in proton fragmentation the diquark 
fragmentation plays an important role, this diquark part of distribution should be 
written separately. So the distribution for $pp$ collision will include 
two diquark parts for positive $x$ as well as for negative:

\begin{eqnarray}
\varphi _{n}^{\Lambda_c}(s,x)=a_f^{\Lambda_c}F_{1qq}^{(n)}(x_{+})+
a_f^{\Lambda_c}F_{1qq}^{(n)}(x_{-})+
a_{0}^{\bar{\Lambda}_c}[F_{q}^{(n)}(x_{+})F_{0qq}^{(n)}(x_{-})+ \\
+F_{0qq}^{(n)}(x_{+})F_{q}^{(n)}(x_{-})+ 
2(n-1)F_{q_{sea}}^{(n)}(x_{+})F_{\bar{q}_{sea}}^{(n)}(x_{-})],\nonumber 
\end{eqnarray}
where $F_{1qq}^{(n)}(x_{+})$ is the distribution at the leading fragmentation
of diquarks, while $F_{0qq}^{(n)}(x_{+})$ is the ordinary part of fragmentation
written with the central density parameter $a_{0}^{\bar{\Lambda}_c}$.
Obviously, the distribution for $\bar{\Lambda}_c$ does not include the leading 
fragmentation term. 

\begin{figure}[ht]
\includegraphics{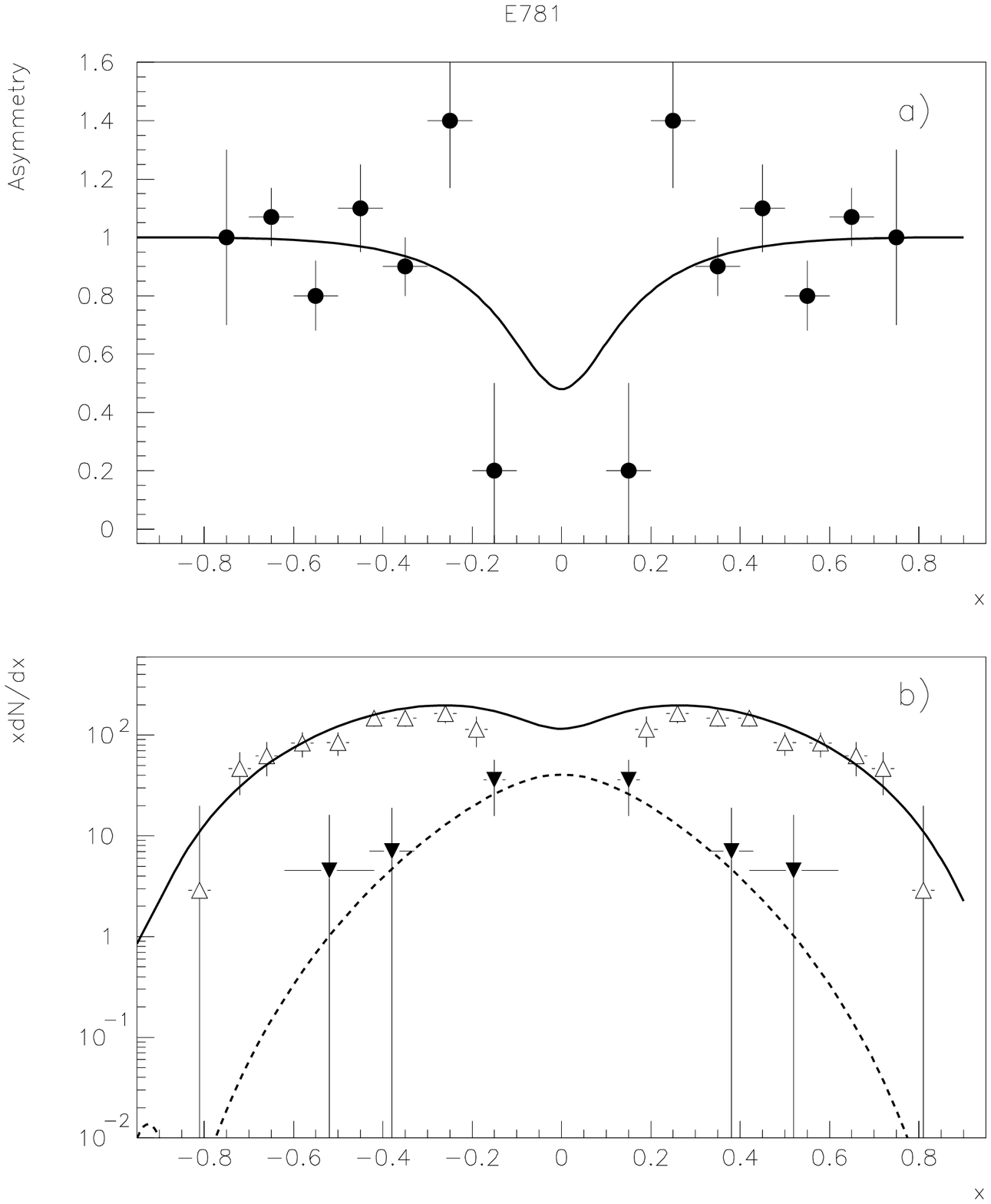}
\caption{\label{FIGURE1} The asymmetry between $\Lambda_c$ and 
$\bar{\Lambda}_c$ spectra obtained for $pA$ collisions 
in the E781 experiment (black circles) \protect\cite{selex},
the QGSM calculation (solid line) ($a$); the distributions of $\Lambda_c$ 
(empty triangles) and $\bar{\Lambda}_c$ (black triangles) in E781 for these reactions 
and QGSM curves: $\Lambda_c$ (solid line) and $\bar{\Lambda}_c$ 
(dashed line) ($b$).}
\end{figure}

The fragmentation functions of diquark and quark chains into charmed baryons 
or antibaryons are based on the rules written in \cite{kaidalov}.
The structure functions of quarks in interacting proton have already been 
described in the previous papers \cite{oldcharm, sigmabeam, pionbeam}. 
The asymmetry between the spectra of $\Lambda_c$ and $\bar{\Lambda}_c$ measured in
$pA$ collisions at $p_L$= 600 GeV/$c$ \cite{selex} is shown in Fig.\ref{FIGURE1}$a$.

The asymmetry is defined as
\begin{equation}
\label{eq:0}
A(x)=\frac{dN^{\Lambda_c}/dx-dN^{\bar{\Lambda}_c}
/dx}{dN^{\Lambda_c}/dx+dN^{\bar{\Lambda}_c}/dx}.
\end{equation}
Here, $dN^{\Lambda_c}/dx$ and $dN^{\bar{\Lambda}_c}/dx$ 
are the event distributions measured in the experiment \cite{selex}. The asymmetry plot has 
the sharp dip in the central region that is peculiar for baryon production 
in proton--proton collisions.

The invariant distributions $xdN/dx$ of charmed baryons and antibaryons obtained in
proton interactions in E781 experiment are shown in Fig.\ref{FIGURE1}$b$ with the QGSM curves 
calculated for proton fragmentation on both sides. It should be mentioned here that usually 
cosmic protons are interacting with the air nuclei in cosmic-ray physics and spectra are
different. But the dependence on atomic number of target nuclei is important at $y<0$ and,
besides, it should be cancelled in formula (\ref{eq:0}). Thus, our conclusions about asymmetries
are valid for proton--nucleus collisions as well.

\section{Charmed baryon/antibaryon spectra in laboratory system}

\begin{figure}[ht]
\includegraphics{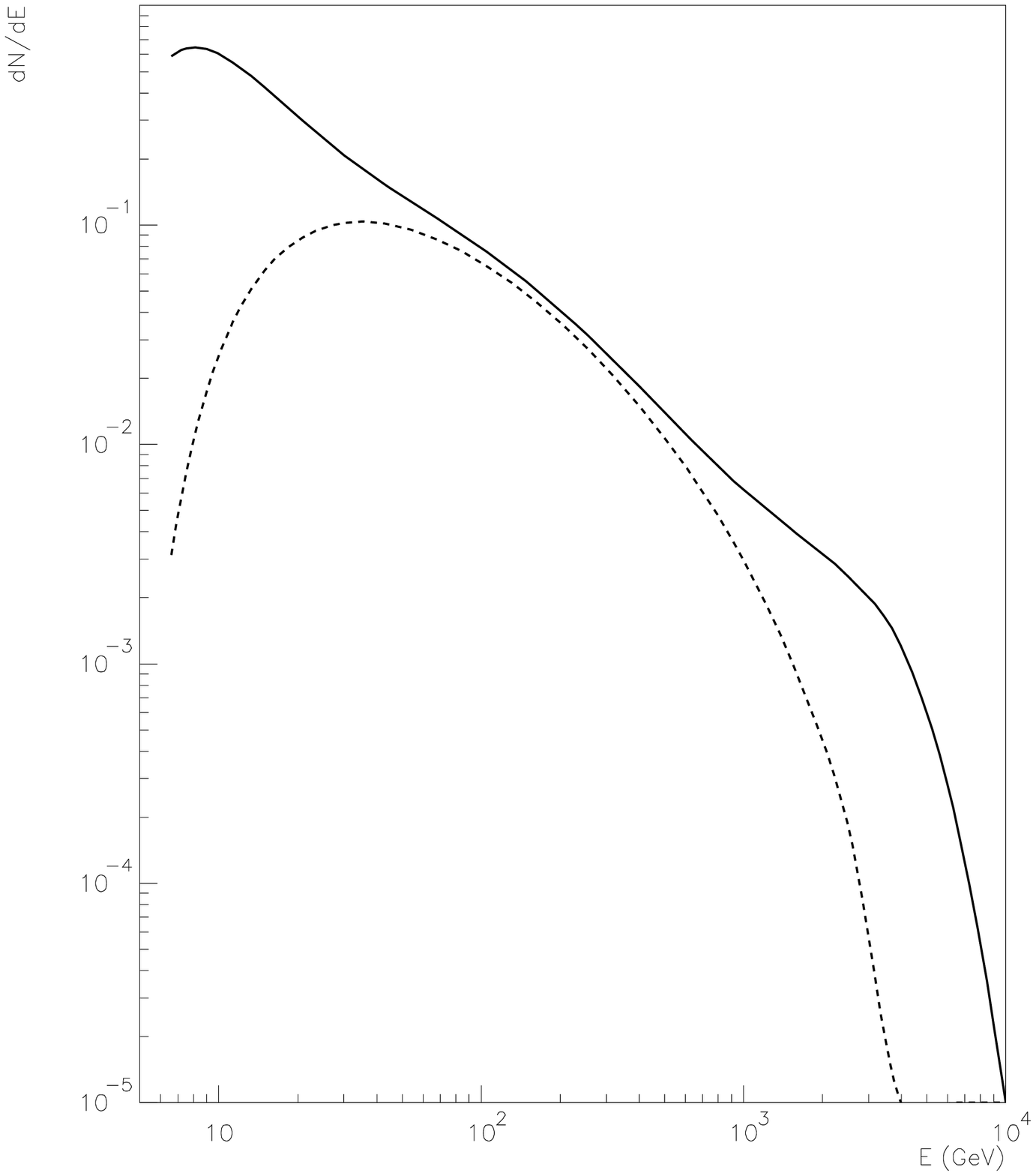}
\caption{\label{FIGURE2} The  distributions of $\Lambda_c$ (solid line) and 
$\bar{\Lambda}_c$ (dashed line) laboratory system, calculated in QGSM for the 
energy $E_p=10^4$ GeV.}
\end{figure}

The results of accelerator experiment are presented usually in center of mass system (c.m.)
that is not accepted in cosmic-ray physics where the Earth is the only possible 
laboratory system (lab.) for the measurements. The transformation of spectra at the 
transition from c.m. to lab. can be done taking into account the invariance 
of the value $d\sigma/dy=xd\sigma/dx_{c.m.}=d\sigma/dy(y_{lab}-y_0)_{lab}$. 
In this case we will have the $E^{-1}$ power slope of the spectra in the laboratory system:

\begin{equation} 
d\sigma/dE=1/E(xd\sigma/dx)=1/E(d\sigma/dy).
\end{equation}

This slope should be seen in the energy region corresponding to the central plateau 
of  distribution in c.m., where $d\sigma/dy=\mbox{const}$. The transformed baryon and antibaryon 
spectra are shown in Fig. \ref{FIGURE2} at the energy of proton interaction $E_p=10^4$ GeV.

This method was already used in the calculation of photon spectra from 
monochromatic cosmic proton sorce \cite{photons} and in the estimation  of antiproton/proton 
ratio \cite{antiprotons} in cosmic rays. In Fig. \ref{FIGURE2} we can see also how the asymmetry 
between spectra behaves in laboratory system. The dip in asymmetry is extending up to the 
energy of order of $0.1 E_p$. Will it be seen in the muon spectra? 

\section{Spectra of muons}

In the previous section we have described the calculations of charmed baryons in 
laboratory system. Here we are giving the brief description of the procedure of 
the calculation of muon spectra that are generated in semileptonic decays of $\Lambda_c$.

It is well known that the rather good approximation to the lepton spectra can be 
obtained for semileptonic decays of charmed baryons, if one takes the probability 
of these decays the same as for exclusive parton decay $c\,\to\, s\mu^+\nu_{\mu}$, where
$c$ quark posseses the energy and the spectrum of $\Lambda_c$.

The decay $c\,\to\, s\mu^+\nu_{\mu}$ has been studied , for example, in \cite{charmdecay}. 
Its differential width is equal to:
\begin{equation}
\label{eq:1}
\frac{d^2\Gamma}{d\hat s\, d\hat t}\, =\, 
\frac{G^2_F}{16\pi^3}\, m_c^5 \left |V_{cs}\right |^2
\left (1+\hat m^2_{\mu}-\hat t \right )\, 
\left (\hat t-\hat m^2_s \right ) ,  
\end{equation}

where $\hat s, \hat t$ are the standart Mandelstam variables, $m_c,m_{\mu}$
are masses of quark and muon. Mandelstam variables $\hat s, \hat t$ have to satisfy
the followig kinematical restrictions:
\begin{equation}
\label{eq:2}
\hat m^2_{\mu}\,\le\,\hat s\,\le\,\left (1-\sqrt{\hat m^2_s} \right ) , 
\qquad
\hat m^2_s\,\le\,\hat t\,\le\,\left (1-\sqrt{\hat m^2_{\mu}} \right ). 
\end{equation}

Let us write formula (\ref{eq:1}) as the expression normalized to the interval 0 to 1
that will be usefull at the construction of Monte-Carlo (MC) generator:
\begin{equation}
\label{eq:3}
w(\hat s,  \hat t)\, =\,
\frac{4}{\left ( 1+\hat m^2_{\mu}-\hat m^2_s \right )^2}
\left ( 1+\hat m^2_{\mu}-\hat t \right )\, 
\left (\hat t-\hat m^2_s \right ). 
\end{equation}

The MC procedure of generating of muon spectra consists of a few steps. 
First of all we are considering the decay $c\,\to\, s\mu^+\nu_{\mu}$ in
the rest system of charmed baryon that is also the rest system of $c$ quark 
and getting the random pair $\hat s, \hat t$ from the interval that was 
written in Eq.(\ref{eq:2}). This pair is being checked with the help of 
kinematical function whether it belongs to the physical space of the process 
(right pair). If the pair is chosen as right it possesses the weight of Eq.(\ref{eq:3}).
The momenta and energies of the decay products are calculated due to $s,t$.

The revolting on three axes are applied to resulting vectors of momenta
that are turned with random Eiler angles to reach the arbitrary positions 
toward the axes. It helps us to return to the system where $\Lambda_c$ is
moving along the $x$ axis. We are neglecting here the transverse motion 
because of the small ratio between transverse momentum and longitudinal one.

As the result of described MC procedure, the table of four-momenta of all products 
of decay is built, where the data for further analysis can be found.

\section{Resulting plots and comparisons}

In this work we have the possibility to analyze the $\Lambda_c$ spectra that can be
calculated analytically with QGSM for arbitrary energy and at the same time to compare
these spectra with the muon spectra that are generating after the decay of charmed baryons.
We are interested here in only muon characteristics, but the possibility exists to
analyze the neutrino spectra too.

\begin{figure}[ht]
\includegraphics{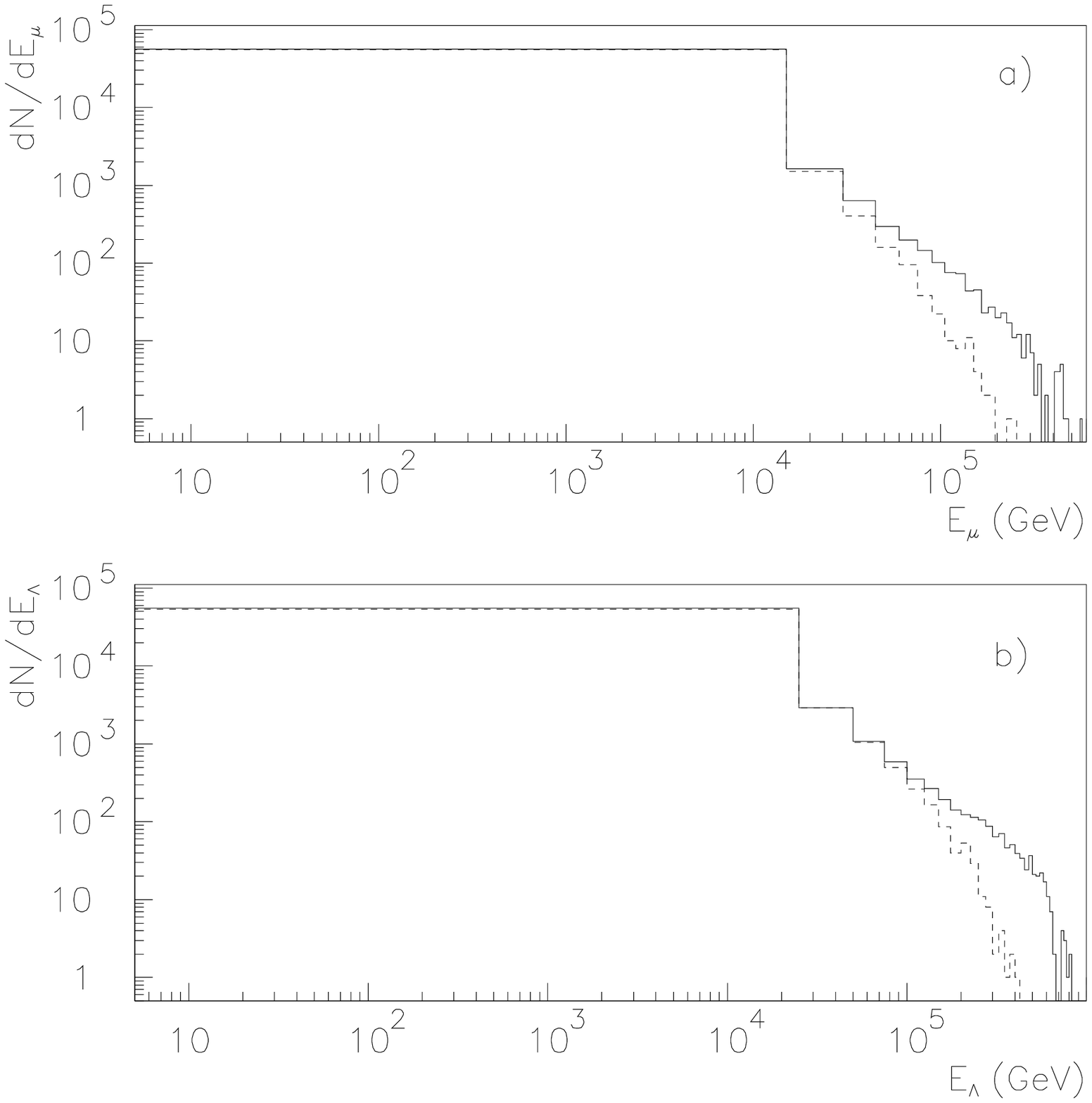}
\caption{\label{Figure3} The spectra of $\mu^+$ (solid line) and $\mu^-$ (dashed line) calculated at 
$E_p=10^5$ GeV ($a$); the distributions of $\Lambda_c$ (solid line) and $\bar{\Lambda}_c$ 
(dashed line) in QGSM for the energy $E_p=10^5$ GeV ($b$).}
\end{figure}

Let us compare the spectra of $\Lambda_c$/$\bar{\Lambda_c}$ with the spectra of
decay products at the energy of initial proton--proton interaction $E_p=10^8$ GeV.
As it can be seen from Fig. \ref{Figure3}, the asymmetry predicted between the spectra of
baryons and antibaryons is clearly reproduced in the spectra of $\mu^+$ and $\mu^-$. 
The spectra of $\Lambda_c$ are of the same form as it was calculated analytically (see Fig. 
\ref{FIGURE2}).

\begin{figure}[ht]
\includegraphics{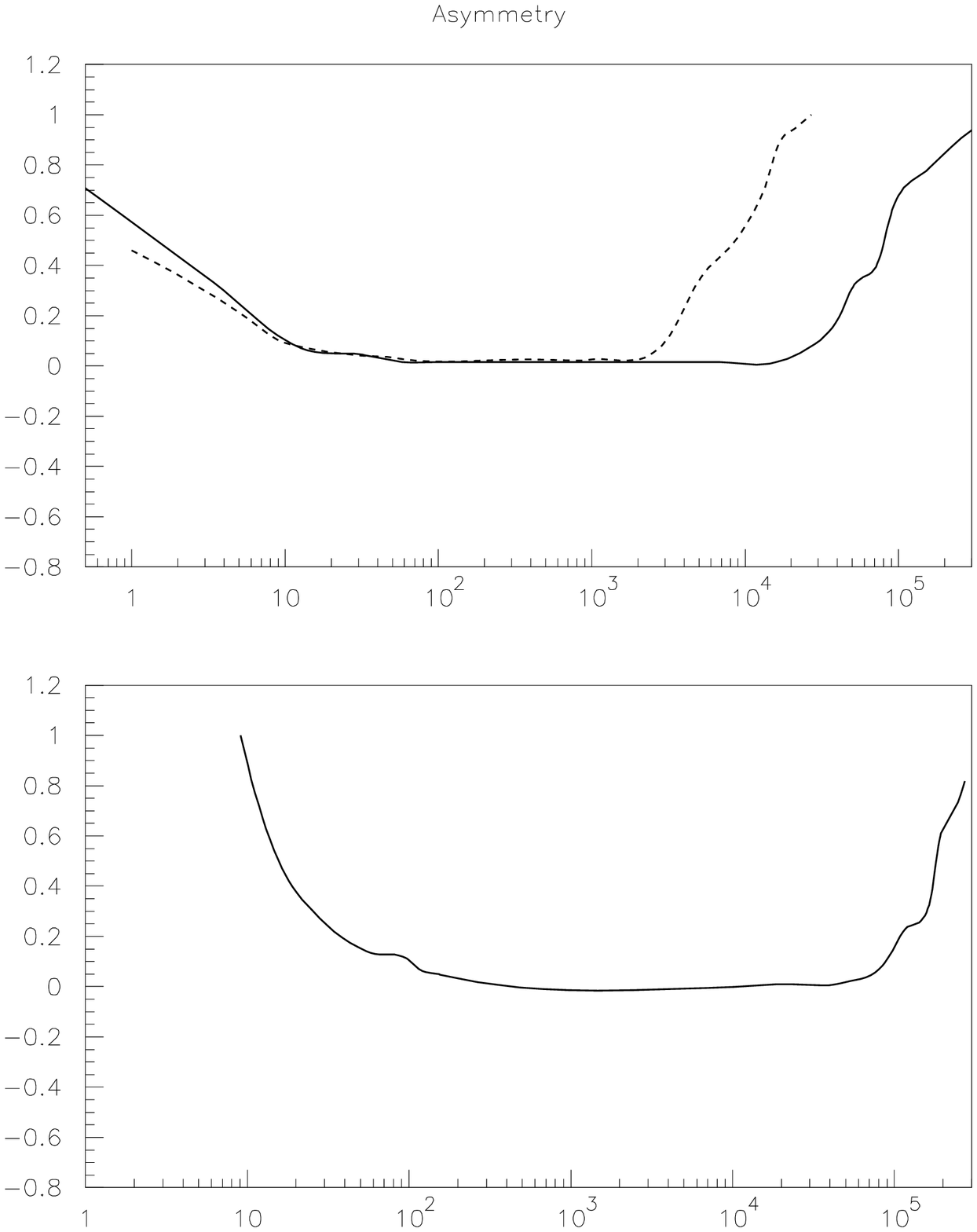}
\caption{\label{Figure4}The asymmetry between spectra of $\mu^+$ and $\mu^-$ calculated at 
$E_p=10^6$ and $10^5$ GeV (dotted line) ($a$); the asymmetry between $\Lambda_c$ and $\bar{\Lambda}_c$ spectra obtained for $pp$ collisions in the 
QGSM calculation at $E_p=10^6$ GeV ($b$).}
\end{figure}

The asymmetry between spectra of $\mu^+$ and $\mu^-$ that is shown in Fig.\ref{Figure4} is almost
equal to zero in the wide region corresponding to the central plateau in $\Lambda_c$-production
distribution. It would be interesting to study how this asymmetry is sensitive to the
difference in baryon/antibaryon production at high energies that might remain to be 
valuable due to the effect of string-junction transfer, wich was not accounted for in our
calculations.

\printfigures

\section{Summary} 

In this paper we have studied the spectra of muons after the semileptonic decay of 
charmed baryons produced in the hadronic interactions at very high energy. It is assumed that the 
conclusions on this research are interesting for cosmic ray physics as well as for high-energy
accelerator physics. 
The asymmetry in production of baryons over antibaryons is caused by the positive
baryonic charge of colliding protons. This asymmetry is to disappear in central region with 
the rising of interaction energy. The spectra of charmed baryons and antibaryons have been
described in Quark--Gluon String Model with the account of quark interaction mechanisms
providing the baryon/antibaryon asymmetry. Only the mechanism of string-junction transfer
was not accounted yet in our scheme. Even so, the obvious asymmetry is reproduced on the edges 
of the spectra of $\Lambda_c$/$\bar{\Lambda_c}$ in laboratory system as it was analytically 
predicted. The knee in the spectra at ~ 0.1 $E_p$ is caused by the form of analytical spectra 
in fragmentation region. This very form of spectra in laboratory system can be a good 
manifestation of the interaction of monochromatic primary protons. 
The Monte-Carlo generator, which was built for the calculation of spectra of products
of charmed baryon decays, provides the four-momenta of all particles after semileptonic decay.
We have analyzed here only muon spectra. These spectra reproduce the production asymmetry 
between positive and negative particles, as it is seen in spectra of the $\Lambda_c$/$\bar{\Lambda_c}$.
We can conclude that muons can be a good instrument to study the baryon production asymmetry in
high-energy proton--proton interactions. The estimation of background from $D$-meson decays should be
done in the same QGSM approach.

\begin{acknowledgments} 
Authors are thankful to Prof. A.B.~Kaidalov and Dr. L.V.~Volkova for the
numerous discussions.One of the authors is supported with the grant of 
RFBR No.04-02-16757a, the grant RFBR-DFG 03-02-04026a and the grant 
NATO CLG.980335.
\end{acknowledgments}


\begin{thebibliography}{99}
\bibitem{cosmic_muons} F.~Halsen {\it et al.}, Phys. Rev. D {\bf 55}, 4475 (1997); 
EAS-TOP Collab., MACRO Collab., Astropart. Phys. {\bf 20}, 641 (2004); K.~Abe {\it et al.}, 
Phys. Lett. B {\bf 564}, 8 (2003); T.~Sanuki {\it et al.}, Phys. Lett. B {\bf 541}, 234 (2002); 
G.~Fiorentini,V. A.~Naumov, F. L.~Villante, Phys. Lett. B {\bf 510}, 173 (2001).
\bibitem{accelerator_charm} R.~Vogt, hep-ph/0412303; in{\it   the Proceedings 
of the Strangenes in Quark Matter, Cape Town, South Africa, 2004}.
\bibitem{general_QGSM}A. B.~Kaidalov and K. A.~Ter-Martirosyan,
Sov. J. Nucl. Phys. {\bf 39}, 1545 (1984); {\bf 40}, 211 (1984);
A. B.~Kaidalov, Phys. Lett. B {\bf 116}, 459 (1982).
\bibitem{previous_QGSM} A. B.~Kaidalov and O. I.~Piskounova, Sov. J. Nucl. Phys.
{\bf 43}, 1545 (1986); O. I.~Piskounova, Phys. At. Nucl. {\bf 56}, 1094 (1993).
\bibitem{kaidalov} A. B.~Kaidalov, Sov. J. Nucl. Phys. {\bf 45}, 1450 (1987).
\bibitem{oldcharm} O. I.~Piskounova, Nucl. Phys. (Proc.Suppl.) B {\bf 93}, 144 (2001).
\bibitem{sigmabeam} O. I.~Piskounova, Phys. At. Nucl. {\bf 64}, 101 (2001).
\bibitem{pionbeam} O. I.~Piskounova, {\it in the Proceedings of the DIS', St.Petersbourg, Russia, April, 2003}.
\bibitem{selex}SELEX Collab.(F. G.~Garcia {\it et al.}), Phys. Lett. B {\bf 528}, 49 (2002).
\bibitem{photons} O. I.~Piskounova, Sov. J. Nucl. Phys. {\bf 51}, 480 (1990).
\bibitem{antiprotons} O. I.~Piskounova, Sov. J. Nucl. Phys. {\bf 47}, 846 (1988).
\bibitem{charmdecay} V.~Novikov {\it et al.}, Phys. Rev. Lett. {\bf 38}, 626 (1977); 
B.~Block, L.~Koyrakh, M.~Shifman, and A.~Vainshtein, Phys. Rev. D {\bf 49}, 3356 (1994). 
\end{thebibliography}
\end{document}